\documentclass[aps,prd,twocolumn,superscriptaddress,preprintnumbers,floatfix,nofootinbib]{revtex4-1}

\usepackage{graphicx}
\usepackage{amsmath}
\usepackage[caption=false]{subfig}
\usepackage{siunitx}
\usepackage{placeins}
\usepackage{color}
\usepackage{standalone}
\usepackage{dcolumn}
\usepackage{tensor}
\usepackage{bm}
\usepackage{microtype}
\usepackage{etoolbox}
\usepackage{amssymb}
\usepackage{mathrsfs}
\usepackage{accents}
\usepackage[normalem]{ulem}
\usepackage[dvipsnames]{xcolor}
\usepackage[colorlinks,urlcolor=NavyBlue,citecolor=NavyBlue,linkcolor=NavyBlue,pdfusetitle]{hyperref}
\usepackage[all]{hypcap}
\usepackage[inline]{enumitem}
\usepackage[utf8]{inputenc}
\usepackage{xstring}
\usepackage{booktabs}

\newcommand{\beq}{\begin{equation}}
\newcommand{\eeq}{\end{equation}}

\newtoggle{commentsoff}
\togglefalse{commentsoff}
\ifdefined\nocomments
    \toggletrue{commentsoff}
\fi

\iftoggle{commentsoff}{
  \newcommand*{\mi}[1]{}
  \newcommand*{\mg}[1]{}
  \newcommand*{\wf}[1]{}
  \newcommand*{\comment}[1]{}
  
  \newcommand*{\todo}[1]{}
  \newcommand*{\warn}[1]{}

  \newcommand*{\commentmark}[1]{}
}{
  \newcommand*{\mi}[1]{{\color{magenta} [{\bf MAX}: #1]}}
  \newcommand*{\mg}[1]{{\color{RedOrange} [{\bf MATT}: #1]}}
  \newcommand*{\wf}[1]{\textcolor{green}{\textbf{WILL}: #1}}
  
  \newcommand*{\commentmark}[1]{{\color{OliveGreen} [{\bf MARK}: #1]}}
  \newcommand*{\comment}[1]{{\color{blue} [{\bf NOTE}: #1]}}
  \newcommand*{\warn}[1]{{\color{red} [{\bf WARNING}: #1]}}
  \newcommand*{\todo}[1]{{\color{red} [{\bf TODO}: #1]}}

}

\graphicspath{{./fig/}}

\newcommand{\dcc}{LIGO-P2000507}

\newcommand{\tevent}{1126259462.423}

\newcommand{\A}{\ensuremath{\mathcal{A}}}
\newcommand{\ProbPos}[1]{\IfEqCase{#1}{{N1}{97\%}{N0}{95\%}}}
\newcommand{\ProbNeg}[1]{\IfEqCase{#1}{{N1}{3\%}{N0}{5\%}}}

\begin{document}

\title{Testing the black-hole area law with GW150914}

\author{Maximiliano Isi}
\email[]{maxisi@mit.edu}
\thanks{NHFP Einstein fellow}
\affiliation{
LIGO Laboratory, Massachusetts Institute of Technology, Cambridge, Massachusetts 02139, USA
}%

\author{Will M. Farr}
\email{will.farr@stonybrook.edu}
\affiliation{Center for Computational Astrophysics, Flatiron Institute, 162 5th Ave, New York, NY 10010}
\affiliation{Department of Physics and Astronomy, Stony Brook University, Stony Brook NY 11794, USA}

\author{Matthew~Giesler}
\affiliation{Cornell Center for Astrophysics and Planetary Science, Cornell
University, Ithaca, New York 14853, USA}

\author{Mark~A.~Scheel}
\affiliation{TAPIR, Walter Burke Institute for Theoretical Physics, California
Institute of Technology, Pasadena, CA 91125, USA}

\author{Saul~A.~Teukolsky}
\affiliation{Cornell Center for Astrophysics and Planetary Science, Cornell
University, Ithaca, New York 14853, USA}
\affiliation{TAPIR, Walter Burke Institute for Theoretical Physics, California
Institute of Technology, Pasadena, CA 91125, USA}

\hypersetup{pdfauthor={Isi, Farr, Giesler, Scheel, Teukolsky}}

\date{\today}

\begin{abstract}
We present observational confirmation of Hawking's black-hole area theorem based on data from GW150914, finding agreement with the prediction with \ProbPos{N1}{} (\ProbPos{N0}{}) probability when we model the ringdown including (excluding) overtones of the quadrupolar mode.
We obtain this result from a new time-domain analysis of the pre- and postmerger data.
We also confirm that the inspiral and ringdown portions of the signal are consistent with the same remnant mass and spin, in agreement with general relativity.
\end{abstract}

\maketitle

\section{Introduction}

The second law of black hole (BH) mechanics, also known as \emph{Hawking's area theorem}, states that the total horizon area of classical BHs cannot decrease over time \cite{Hawking:1971tu}.
This is a fundamental consequence of general relativity (GR) and the cosmic censorship hypothesis \cite{penrose1969,Chrusciel:2000cu}, with far reaching implications for classical and quantum gravity (see \cite{Wald:1999vt} for a review).

If the area theorem is obeyed by binary BH mergers observed by LIGO \cite{TheLIGOScientific:2014jea} and Virgo \cite{TheVirgo:2014hva}, then the combined horizon area of the two progenitor BHs must not exceed that of the remnant BH produced by the merger.
Therefore, gravitational waves (GWs) could provide observational confirmation of Hawking's prediction.
Although this prospect has been discussed in the literature \cite{Hughes:2004vw,Unnikrishnan,Giudice:2016zpa,Cabero:2017avf}, so far no analysis explicitly targeting the BH area has been carried out conclusively on real LIGO-Virgo data.

In this paper, we present observational confirmation of Hawking's area law based on data from LIGO's first detection, GW150914 \cite{Abbott:2016blz}.
We do this by analyzing the inspiral and ringdown portions of the signal independently so as to measure the change in the total horizon area caused by the merger.
We carry out the analysis fully in the time domain, circumventing issues with Fourier frequency mixing and non-periodic boundary conditions \cite{Isi:2019aib,Carullo:2019flw}.
We find the theorem is obeyed with \ProbPos{N1}{} (\ProbPos{N0}{}) probability if we model the ringdown including (excluding) overtones of the quadrupolar mode.
We obtain slightly weaker, albeit consistent, results if we truncate the inspiral at earlier times.

\section{Method}

\newcommand{\Ai}{\ensuremath{\A_0}}
\newcommand{\Af}{\ensuremath{\A_f}}
\newcommand{\NRSur}{\textsc{NRSur7dq4}}

\begin{figure}
\includegraphics[width=\columnwidth]{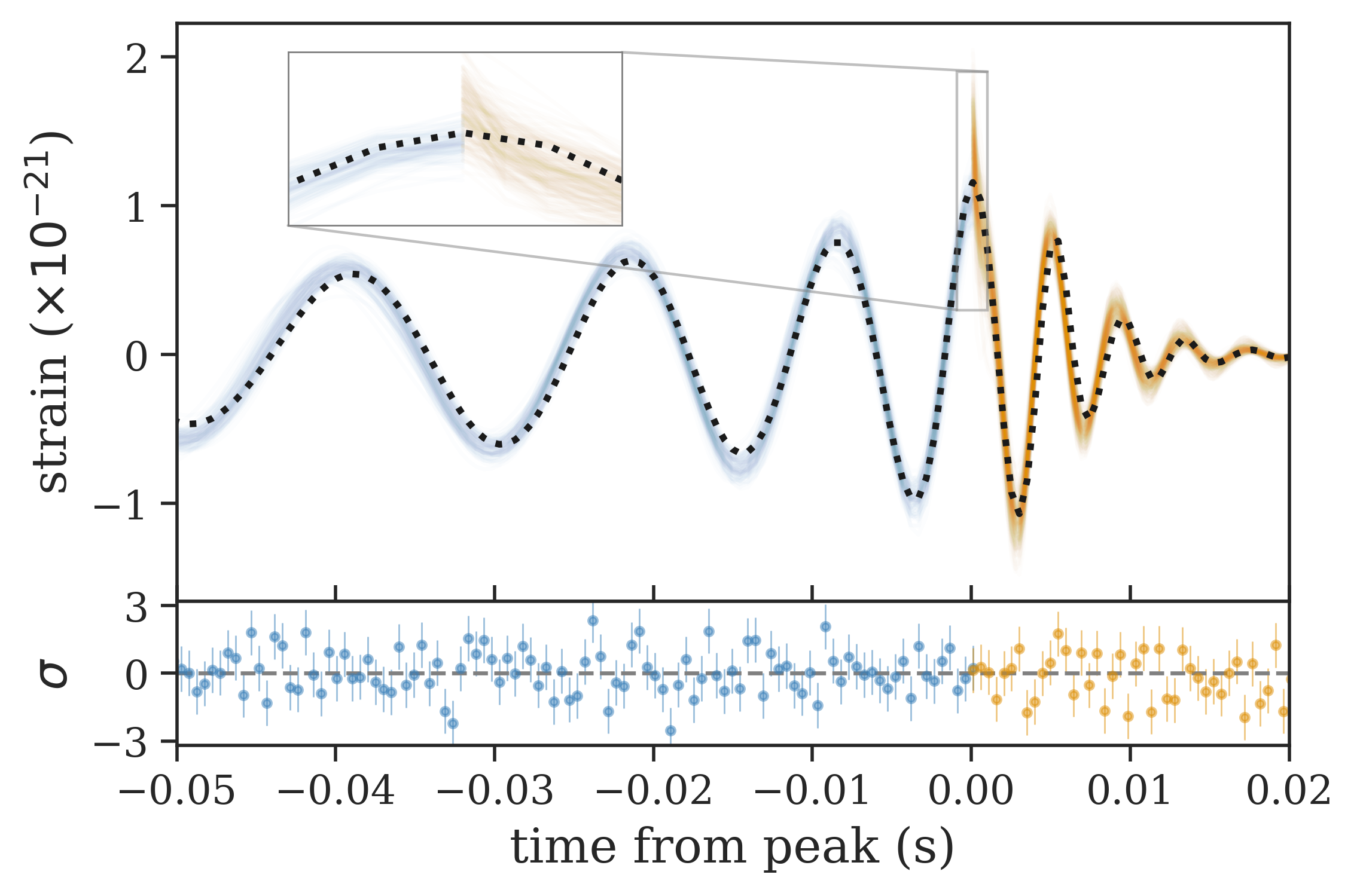}
\caption{GW150914 reconstruction. Hanford waveforms drawn randomly from the posterior of the premerger (blue) and postmerger (orange) analyses, compared to a draw from the full inspiral-merger-ringdown analysis (black). The bottom panel shows the corresponding whitened residuals obtained by subtracting the maximum a-posteriori (MAP) piece-wise waveforms from the data.
The detector data are sampled at 2048 Hz, and the time origin corresponds to the truncation time ($t = \tevent\, \mathrm{s}~\mathrm{GPS}$).
}
\label{fig:strain}
\end{figure}

The horizon area $\A$ of a Kerr BH with mass $M$ and spin angular momentum $\vec{J}$ is
\begin{equation}
\A(M,\,\chi) = 8 \pi \left(\frac{G M}{c^2}\right)^2 \left(1 + \sqrt{1-\chi^2}\right)\, ,
\end{equation}
where $\chi \equiv |\vec{J}|c/(GM^2)$ is the dimensionless spin magnitude.
For two well-separated inspiraling BHs, the total horizon area is simply $\Ai \equiv \A(m_1,\, \chi_1) + \A(m_2,\, \chi_2)$, where $m_{1,2}$ and $\chi_{1,2}$ are the masses and spins of the two components.
The merger produces a remnant BH with mass and spin $m_f$ and $\chi_f$, whose horizon area is $\Af \equiv \A(m_f, \chi_f)$.
Our goal is to independently extract $\Ai$ and $\Af$ from the GW signal in order to compute the change in the total area, $\Delta\A \equiv \Af - \Ai$.

To obtain independent pre- and postmerger measurements, we split the LIGO timeseries data at the inferred peak of the GW signal, and analyze the two resulting segments separately.
We are able to do so by adapting the time-domain Bayesian analysis we developed in \cite{Isi:2019aib} to apply to the inspiral signal, in addition to the ringdown.
For the premerger data, we estimate $m_{1,2}$ and $\vec{\chi}_{1,2}$ using the \NRSur{} waveform model to obtain an accurate representation of the signal up to the peak \cite{Varma:2019csw}.
We place uniform priors on the binary's total mass, mass ratio, spin magnitudes, luminosity volume, and cosine of the inclination, as well as an isotropic prior on the spin orientations; we fix the sky location to the values in \cite{Isi:2019aib}.
We show the resulting reconstruction in Fig.~\ref{fig:strain}.

For the postmerger data, we take advantage of our $m_f$ and $\chi_f$ measurements from \cite{Isi:2019aib}.
In that work, we used BH perturbation theory to infer the remnant parameters from the frequency and damping times of its quasinormal modes, as imprinted on the later portion of the GW150914 signal.
By including overtones in our model \cite{Giesler:2019uxc}, we were able to begin our analysis immediately after the peak of the complex strain---at the same exact point where we have now truncated our inspiral analysis (cf.~Fig.~\ref{fig:strain}).
In \cite{Isi:2019aib}, we also repeated the analysis without overtones but starting at a later time, when we expect only the longest-lived mode to be measurable.
The two types of measurement (multimode at the peak versus single mode after the peak) yielded comparable inferences on $m_f$ and $\chi_f$ (see Fig.~3 in \cite{Isi:2019aib}).
Below we use both measurements, computing $\Af$ based first on a model with one overtone at the peak ($N=1$, $\Delta t_0 = 0~\mathrm{ms}$), and then on one without any overtones 3 ms after the peak ($N=0$, $\Delta t_0 = 3~\mathrm{ms}$), which should be sufficiently late for this signal (e.g., see \cite{Giesler:2019uxc}).
We label the measurements based on the number of overtones included, $N$, and the delay after the inferred peak, $\Delta t_0$.

We contextualize our measurements by comparing them to predictions for the remnant properties based on a coherent analysis of the full inspiral-merger-ringdown (IMR) signal.
As in \cite{Isi:2019aib}, we produce this from the LIGO-Virgo posterior samples released in \cite{LIGOScientific:2018mvr,GWOSC:GWTC1,Abbott:2019ebz}, using numerical-relativity fits to derive $m_f$ and $\chi_f$ \cite{Varma:2018aht,Blackman:2017pcm}.

\section{Results}

\begin{figure}
\includegraphics[width=\columnwidth]{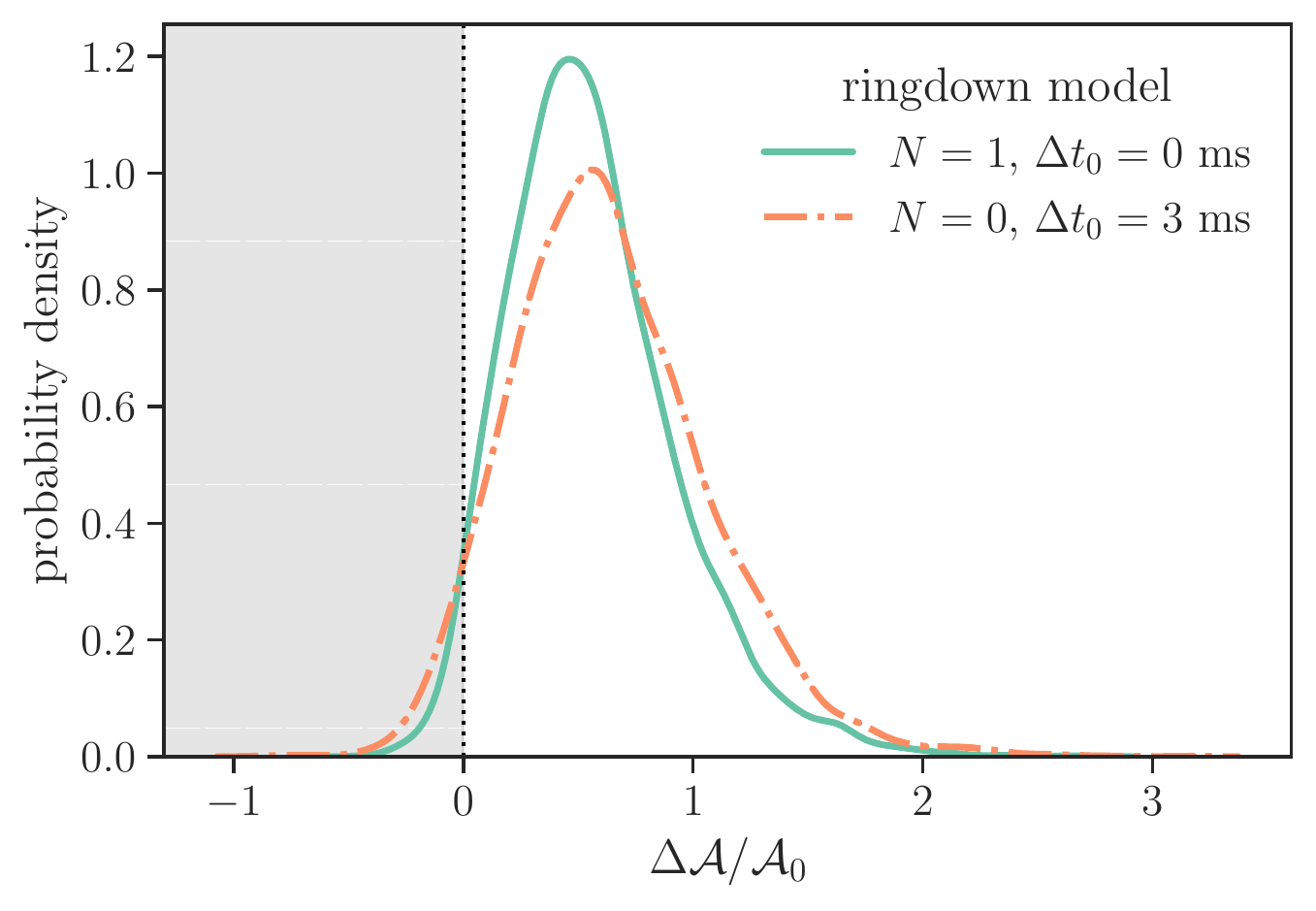}
\caption{Fractional change in the horizon area before and after the GW150914 merger, $\Delta\A/\Ai = \left(\Af - \Ai\right)/\Ai$. We infer the premerger area, $\Ai$, from the inspiral alone (Fig.~\ref{fig:strain}).
We infer the postmerger area, $\Af$, from the remnant mass and spin as estimated from an analysis of the ringdown using the fundamental mode and one overtone at the peak (green), as well as solely the fundamental mode 3 ms after the peak (orange).
For the former (latter), we measure $\Delta\A/\Ai = 0.52^{+0.71}_{-0.47}~(0.60^{+0.82}_{-0.60})$ at 90\% credibility, and find agreement with Hawking's area theorem with \protect\ProbPos{N1}{} (\protect\ProbPos{N0}) probability.
}
\label{fig:area}
\end{figure}

\begin{figure}
\includegraphics[width=\columnwidth]{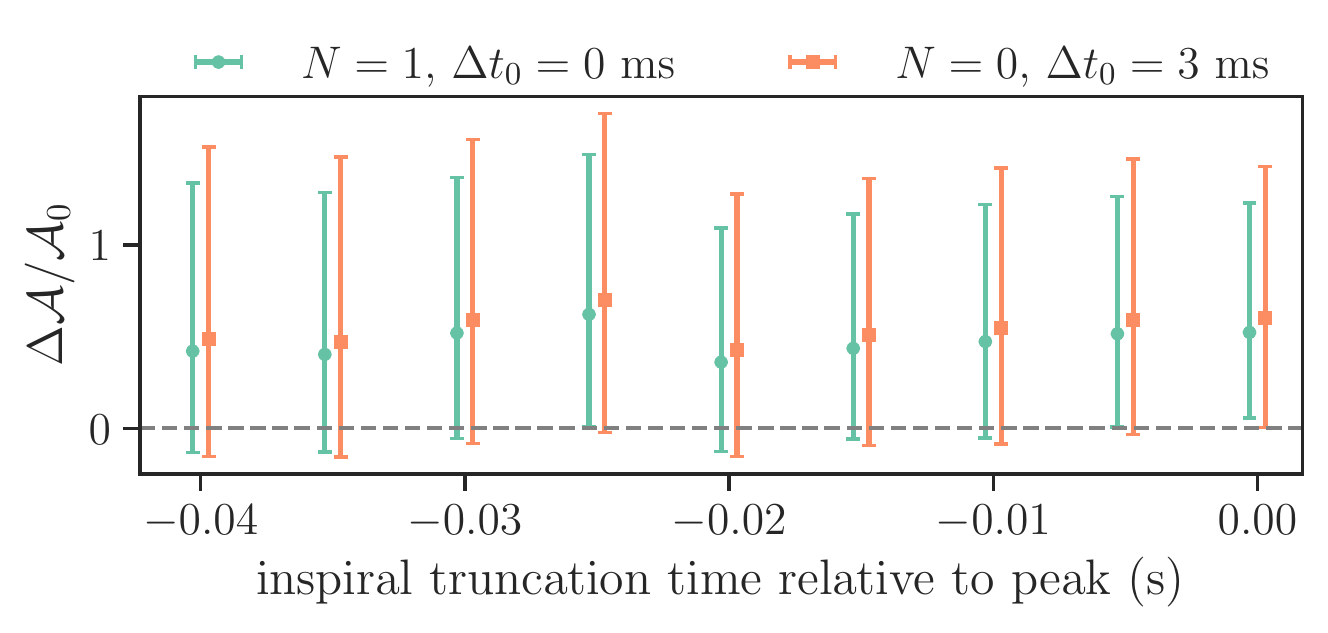}
\caption{Measurements of $\Delta \A/\A_0$ (ordinate) obtained by truncating the inspiral at different times before the peak (abscissa), and with respect to two ringdown measurements (color).
Bars show the symmetric 90\%-credible interval around the median, itself indicated by a marker.
The rightmost points correspond to the distributions in Fig.~\ref{fig:area}.
All measurements support the area theorem, with probabilities within $88{-}97\%$.
}
\label{fig:insp_trunc}
\end{figure}

Figure~\ref{fig:area} summarizes the main result of our analysis.
Whether we infer the remnant parameters with two modes at the peak (green) or a single mode 3 ms after the peak (orange), our measurement favors $\Delta \A \geq 0$, in agreement with Hawking's area theorem.
We can assert that $\Delta \A \geq 0$ with \ProbPos{N1}{} credibility if relying on the overtone, or \ProbPos{N0}{} if not.
The second measurement is less constraining because of the rapid decay of the signal after peak amplitude \cite{Isi:2019aib}.

We check the robustness of our analysis by truncating the inspiral at progressively earlier times.
This leads to slightly weaker but consistent results, showing agreement with Hawking's theorem even for truncation times significantly before the peak (Fig.~\ref{fig:insp_trunc}).
All measurements confidently imply $\Delta \A < 3 \Ai$, as would be required by conservation of energy ($m_f < m_1 + m_2$) \cite{Giudice:2016zpa}.

\begin{figure}
\includegraphics[width=\columnwidth]{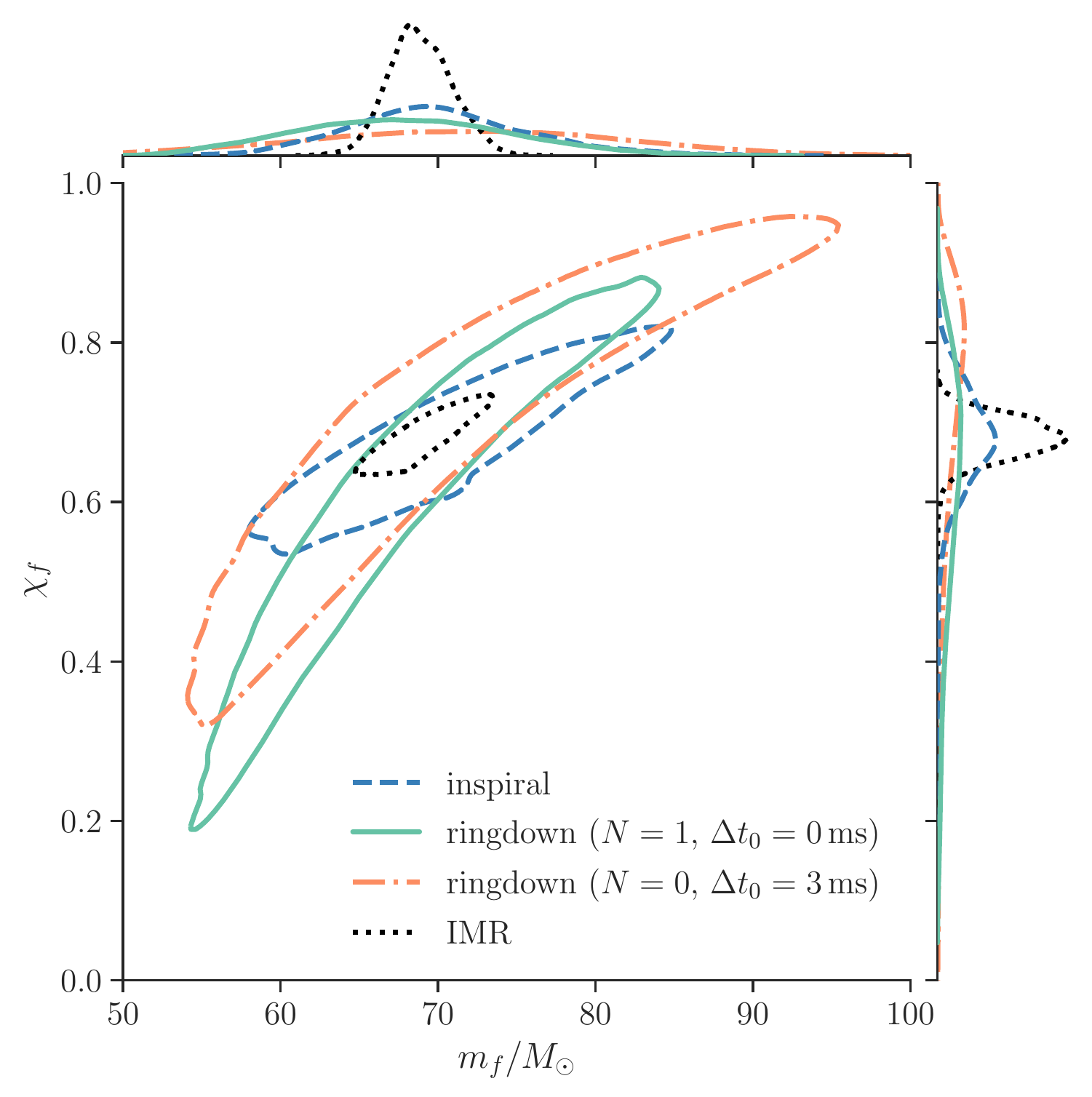}
\caption{Redshifted remnant BH mass $m_f$ (abscissa) and dimensionless spin $\chi_f$ (ordinate), as inferred from different segments of the GW150914 data.
One measurement is based on the prepeak inspiral data alone (dashed blue); two others focus on the postpeak ringdown data,  either using the fundamental mode plus an overtone at the peak (solid green) or just the fundamental 3 ms after the peak (dash-dotted orange); one final measurement relies on the full inspiral-merger-ringdown signal (dotted black).
Contours enclose 90\% of the probability mass, while the top and right panels show the $m_f$ and $\chi_f$ marginals respectively.
All measurements agree.
}
\label{fig:remnant}
\end{figure}

The independent pre- and postmerger measurements can also be used to more broadly evaluate the consistency of the signal with the prediction from GR.
In Fig.~\ref{fig:remnant} we do this by comparing the properties of the remnant as inferred from the different portions of the signal, as is regularly done for LIGO-Virgo data \cite{Ghosh:2016qgn,Ghosh:2017gfp,TheLIGOScientific:2016src,LIGOScientific:2019fpa,Abbott:2020jks}: if GR is valid and the signal was produced by Kerr BHs, the different measurements should all be consistent with some overlapping set of $m_f$ and $\chi_f$ values.
This is the case in Fig.~\ref{fig:remnant}, which shows that the 90\%-credible contours for the inspiral (blue) and ringdown (green and orange) measurements all agree with the each other, as well as with the result from analyzing the full IMR signal coherently (black);
Table \ref{tab:remnant} shows the corresponding  90\%-credible measurements for the individual parameters.

\section{Discussion}

Figure \ref{fig:area} shows that the GW150914 data highly support agreement with Hawking's theorem, whether we extract the properties of the remnant starting at peak strain with an overtone ($N=1$, $\Delta t_0 = 0$) or at a later time without it ($N=0$, $\Delta t_0 = 3\, \mathrm{ms}$).
Although the measurement at the peak is slightly more constraining, it is computed under the assumption that BH perturbation theory can offer a complete description of the data starting right at the peak.
This expectation is based on recent studies of numerical relativity simulations for nonprecessing systems, with particular focus on a high-accuracy numerical simulation of a GW150914-like system \cite{Giesler:2019uxc}.
Exploring the extent of overtone models beyond nonprecessing systems, the resolvability of overtones in data analysis, and the apparent lack of non-linearities in binary black hole mergers remain active research topics \cite{Bhagwat:2019dtm,Ota:2019bzl,Okounkova:2020vwu}.
In this respect, the measurement using only the fundamental mode serves as a more conservative approach.

\begin{table}
\caption{90\%-credible measurements of $m_f$ and $\chi_f$ (Fig.~\ref{fig:remnant}).}
\label{tab:remnant}
\begin{tabular}{l@{\qquad}l@{\qquad}l}
\toprule
{} &      $m_f/M_\odot$ &                $\chi_f$ \\
\midrule
inspiral &   $69.58^{+10.74}_{-8.52}$ &  $0.68^{+0.10}_{-0.12}$ \\[2pt]
ringdown ($N=1,\, \Delta t_0 =0\,{\rm ms}$) &   $67.64^{+11.76}_{-10.71}$ &  $0.63^{+0.19}_{-0.35}$ \\[2pt]
ringdown ($N=0,\, \Delta t_0 =3\,{\rm ms}$) &   $71.64^{+16.00}_{-16.82}$ &  $0.74^{+0.16}_{-0.40}$ \\[2pt]
IMR   &   $68.77^{+3.57}_{-3.05}$ &  $0.68^{+0.04}_{-0.04}$ \\[2pt]
\bottomrule
\end{tabular}
\end{table}

A caveat to our analysis lies in the choice of truncation time, which is itself informed by a GR-based reconstruction of the IMR signal, and is affected by statistical noise.
This means that our chosen truncation time may not exactly agree with the true signal peak.
However, for waveforms reasonably close to GR, we should expect the corresponding posterior error to be smaller than the statistical uncertainty.
Again, the $N=0$ measurement is more robust in this respect thanks to the 3 ms gap of buffer data after the peak.
Similarly, the shortened-inspiral measurements in Fig.~\ref{fig:insp_trunc} are also more conservative.

The consistency test based on the properties of the remnant (Fig.~\ref{fig:remnant}) is comparable to previous analyses in \cite{Ghosh:2016qgn,Ghosh:2017gfp,TheLIGOScientific:2016src,LIGOScientific:2019fpa,Abbott:2020jks}.
However, it is novel in being implemented fully in the time domain, for both the pre- and postmerger measurements.
Working in the time domain allows for a better-defined separation between the two regimes, without risk of being affected by Fourier frequency mixing.
It also allows us to apply a postmerger model manifestly based on perturbation theory alone, without relying on phenomenological waveform approximants that could suffer from modeling systematics.

\section{Conclusion}
We have confirmed that the GW150914 data agree with Hawking's area theorem with high probability (${>}95\%$ or ${\sim}2\sigma$).
This result stems from separately analyzing the data before and after the merger, which can also be used to carry out a GR consistency test in the space of remnant parameters ($m_f$, $\chi_f$).
Our measurements further demonstrate the potential of time-domain analyses of LIGO-Virgo data, and pave the way for more stringent tests of Einstein's theory with future GW detections.

\begin{acknowledgments}
We thank Kip S.\ Thorne for encouraging this study, and providing feedback during its development;
we also thank Geraint Pratten, Nathan Johnson-McDaniel and Katerina Chatziioannou for comments on the draft.
M.I.\ is supported by NASA through the NASA Hubble Fellowship
grant No.\ HST-HF2-51410.001-A awarded by the Space Telescope
Science Institute, which is operated by the Association of Universities
for Research in Astronomy, Inc., for NASA, under contract NAS5-26555.
The Flatiron Institute is supported by the Simons Foundation.
M.G. and S.T.\ are supported in part by the Sherman Fairchild
Foundation and by NSF Grant PHY-1912081 at Cornell.
M.S.\ is supported in part by the Sherman Fairchild Foundation and NSF
grants PHY-2011961 and PHY-2011968 at Caltech.
This research has made use of data, software and/or web tools obtained from the Gravitational Wave Open Science Center (https://www.gw-openscience.org/), a service of LIGO Laboratory, the LIGO Scientific Collaboration and the Virgo Collaboration. LIGO Laboratory and Advanced LIGO are funded by the United States National Science Foundation (NSF) as well as the Science and Technology Facilities Council (STFC) of the United Kingdom, the Max-Planck-Society (MPS), and the State of Niedersachsen/Germany for support of the construction of Advanced LIGO and construction and operation of the GEO600 detector. Additional support for Advanced LIGO was provided by the Australian Research Council. Virgo is funded, through the European Gravitational Observatory (EGO), by the French Centre National de Recherche Scientifique (CNRS), the Italian Istituto Nazionale della Fisica Nucleare (INFN) and the Dutch Nikhef, with contributions by institutions from Belgium, Germany, Greece, Hungary, Ireland, Japan, Monaco, Poland, Portugal, Spain.
The authors are grateful for computational resources provided by the LIGO Laboratory and Cardiff University, and supported by NSF Grants PHY-0757058 and PHY-0823459, and STFC Grant ST/I006285/1.
This paper carries LIGO document number \dcc{}.
\end{acknowledgments}

\bibliography{refs}

\end{document}